\documentclass[12pt]{article}

\usepackage{amsmath,amsthm,amssymb,color}
\usepackage{epsfig,amsfonts,latexsym, hyperref, bbm, mathrsfs}
\begin{document}	

\begin{center}
	{\bf\large{Particle on a Torus Knot: Anholonomy and  Hannay Angle }}\\
	\vspace {3cm}
	{\bf{Subir Ghosh}}\\
	
	\vspace {1.5cm}
	Physics and Applied Mathematics Unit,\\
Indian Statistical Institute,\\
203 B. T. Road, Kolkata 700108, India.
		
\end{center}
	
	\vspace{1cm}
	{\bf{Abstract:}} The phenomenon of rotation of a vector under parallel transport along a closed path is known as anholonomy. 	In this paper we have studied the anholonomy for noncontractible loops - closed paths in a curved surface that do not  enclose any area and hence Stokes theorem is not directly applicable. Examples of such closed paths are poloidal and toroidal loops and knots on a torus. The present study  is distinct from conventional results on anholonomy for closed paths on $S_2$ since in the latter case all closed paths are contractible or trivial cycles. We find that  for some nontrivial cycles the anholonomy cancels out over the complete cycle. Next we  calculate Hannay angle  for a particle traversing such noncontractible loops when the torus itself is revolving. Some new and interesting results are obtained especially for poloidal paths that is for paths that encircle the torus ring.
	\vskip 1cm
{\bf{Introduction:}}\\
Over the years the boundary between physics and mathematics is getting more and more diffused and  both disciplines are  being enriched by their mutual overlap. 
Knot theory, our present interest,  has started to impact deeply the physics community after  the seminal work of   Jones \cite{jones}, eg. Jones polynomial (related to knot invariants). Subsequent works of Witten \cite{wit}, on one hand, established its connection with topological quantum field theory, wheras, on the other hand, via Temperley-Lieb  relations \cite{temp}   direct connections with statistical mechanical models (such as  $Q$-state Potts model) was defined. Now  knot theory is playing significant role even in  mathematical
biology since it has been established experimentally that 
certain types of DNA molecules take
the form of certain types of knots. Coming back to physics, knotted structures 
in physical fields appeared first in the hypothesis of  Lord Kelvin's vortex atom model. The first concrete indication of knot-like structure 
in physics appeared in the work of  Faddeev and Niemi \cite{fadd} who identified such structures with  3+1 dimensional stable finite energy solitons and suggested that similar structures might be observed in nematic liquid crystals and $He^3$ superfluids.  Only recently they are visible  experimentally in diverse  physical systems, such as in fluid vortex
lines, singular lines of optical fields, the topological defect lines in liquid
crystals,  magnetic
field lines in electromagnetic fields, and in
spinor Bose-Einstein condensates (we quote from \cite{prl} where references are provided).  Probably one of the simplest and most relevant example of knot  is the torus knot. It is a   knot that lies on the surface of an unknotted torus in three dimensional space. A beautiful visualization of  general $(p, q)$-torus knots can be found in \cite{prl} where a family of exact knotted solutions to Maxwell’s equations in free
space are represented as torus knots and links constructed out of the   electric and magnetic field lines that,  surprisingly, persist for all time. 

Although this is not directly along the lines of our paper still it is important to realize the importance of toroidal geometry in physics and compactified quantum field theory on a torus (for a review and exhaustive list of references see \cite{mal}). Indeed, it has become extremely relevant in recent times because of the growing interest in finite size effects 
especially due to the vastly improved experimental scenario. The significance of  toroidal topology is  that the local properties of the space-time remain unchanged and the non-trivial   topological effect   manifests, for example, in  correlation
functions, which are  non-local properties of the system.

It is now time to be more specific and concentrate on the problem at hand. In the present paper we consider a different kind of influence of knot   on physics where a particle is constrained to move on a torus knot embedded in Euclidean three space. This type of generic toy model of particles moving on a predetermined path (of various levels of  complexity) actually helps in understanding deeper issues, such as existence of
inequivalent quantizations of a given classical system \cite{n1}, the role of topology in the definition of the vacuum state in
gauge theories \cite{n2}, band structure of solids \cite{n3}, generalized spin and statistics of the anyonic type \cite{n4}, and the study of
mathematically interesting algebras of quantum observables on spaces with non-trivial topology \cite{n5}, to name a few.
Since a  particle moving on a circle is identified to a (single) rotor, a particle on a torus, considered here, can be identified to a 
 double-rotor, acting as a non-planar extension  of the planar
rotor \cite{sree}.

In the present paper we will address two problems: Anholonomy associated with torus knot paths and Hannay's angle referring to such paths.

{\bf{Anholonomy:}} Nonintegrability introduces the geometrical phenomenon of anholonomy when some parameters do not return to their original values even when other quantities that drive the said parameters come back to their original values after a complete cycle. The classic example is the rotation of a vector that has been Parallel Transported (PT) along a closed path on a curved surface. 

{\bf{(Nonintegrable) Hannay angle:}}
It was discovered by Berry  \cite{b1} that a similar effect is present in a quantum system where the anholonomy in parameter space manifests itself as a geometrical phase (the Berry phase), apart from the conventional dynamical phase  the wave function accumulates while evolving along a closed path in parameter space. It is interesting to observe that the idea of Hannay angle \cite{h1} in classical physics emerged from  the notion of Berry's phase in quantum physics and in fact the former is obtained from the latter in the classical limit. A precursor of these ideas is the early works of Pancharatnam \cite{panch} on light polarization and the Aharanov-Bohm effect \cite{ab}. 

The Hannay angle \cite{h1} in classical mechanics, a geometric effect in systems exhibiting periodic motion, can appear as an extra revolution in angle variables  that can be  physical (configuration space)  angles or more generally  angle variables in abstract action-angle  framework.  Furthermore,  it was shown explicitly that  Hannay angle in classical scenario will yield Berry phase in quantal scenario since  the former appears as a semi-classical limit of the latter \cite{b3}. ( For discussions on Berry phase without a Hannay angle see \cite{gozz}).

   Applications of Hannay angle range from rotation of the plane of oscillation in Foucault's pendulum \cite{fou}, generalized oscillators with time-dependent frequency \cite{song}, time-dependent Hamiltonian systems consisting of
 several degrees of freedom \cite{golin} to celestial dynamics \cite{gol,sp1,mor,sp2}. In \cite{gol} Hannay angle was exploited in  measurements in the solar system related to  non-sphericity
and slow rotation of the Earth.  In \cite{sp1} Spallicci et.al. have generalized and improved upon  an earlier work \cite{mor} on Hannay angle for restricted celestial three-body system in circular motion using adiabatic approach. In \cite{sp2} an interesting experiment is proposed using an Earth-bound
satellite, adiabatically driven by the Moon and explicit estimates show that it is indeed possible to evaluate Hannay angle in high Earth orbits, using atomic clocks, accurate time and frequency
transfer system and precise positioning{\footnote{ We quote from \cite{sp2} numerical estimates of  Hannay angle of a geostationary satellite perturbed by the Moon. It ia around $6.204 . 10^{-8}$ 		radians for
		a perturber period. The values suggest (for details see \cite{sp2}) that the best  chances of observations are for geosynchronous satellite orbits.}}. The Hannay angle
appears \cite{sp1,sp2} as a precession and it is identified with the forward displacement of the perturbed body on the trajectory at a given orbit radius. Hannay and Berry \cite{chaos} explored a model of quantum mechanics on the torus in order to study   quantum
chaos in this model. Following their suggestion of  simultaneous quantization of the functions on
the torus and the linear symplectic group $Sp(2, Z) = SL(2, Z)$ the authors of \cite{had} have considered quantizing the full symplectic group for two and higher dimensional tori. In another theme, interplay between fast (variation of the angular momentum responsible for inertia) and slow time scale variables (ferromagnetic moment) appear in \cite{spie} via Hannay angle in ferromagnetic sample in the precession of the magnetization in the case of the inertial effect, and the corresponding magnetic monopole. Very recently classical and quantum dynamics of particle on torus has generated a lot of interest \cite{sree,g1}. (Earlier important works where particle dynamics on torus is relevant are cited in \cite{sree}.)

Conventionally Berry phase or Hannay angle are expressed in terms of area (or solid angle) associated with the closed path since {\it{in generic examples the closed loops or  cycles    are taken to be bounding cycles}} meaning that they enclose an area in ${\bf{R}}$ (or in a surface embedded in ${\bf{R}}$).  Subsequently Stokes law is applied to introduce the area enclosed by the loop to finally reproduce the well known results in question, {\it{i.e.}} anholonomy, Berry Phase or Hannay angle. In standard examples one considers the parameter space to be  $S^2$ in ${\bf{R}}$ on which all closed loops are bounding. We will always talk about non-self-intersecting loops{\footnote{I thank Professor Berry for pointing this out.}}.

	The question that immediately comes to mind, (and will be dealt with in the present paper), is what will happen if {\it{the closed paths are non-bounding cycles that do not enclose any area}}.  Surfaces having "handles" allows such possibility, torus being the simplest example. Toroidal (fixed $\theta$ in (\ref{1})) below and poloidal  (fixed $\phi$ in (\ref{1})) loops  (see Figure 1) or a combination of both in the form of knot or unknot  are examples of noncontractible cycles. In this paper, in the first part, we will compute anholonomy related to this type of loop. Then in the second part, we will move on to the dynamical problem of a particle  traversing these cycles on a {\it{revolving}} torus and compute  the associated   Hannay angle.   Our main finding is that the essence of non-boundedness is captured by the poloidal loops and results connected to them are novel (and not obvious) whereas the toroidal loop results are essentially extensions of the planar hoop model \cite{h1}.

	The paper is organized as follows: 
		To get a feel of calculus on a torus, first  we compute anholonomies for non-bounding cycles on a torus. After this  we come to the main topic of interest: Hannay angle \cite{h1}  connected to particle dynamics on a revolving torus loop from two different frameworks \cite{b3}). 	 We will conclude with mentioning directions for further study. Throughout the paper our approach and presentation will be inclined more towards physics oriented applications.
	\vskip .3cm
	 {\bf{Parallel Transport along noncontractible loops on torus and anholonomy}}\\
	 Parallel Transport (PT)  is a prescription to compare two  vectors in curved manifold at two different locations. PT'ing a vector along a closed path in a curved manifold can change its orientation whereas the same in a flat manifold will reproduce the same vector after traversal of a closed path. In a physically intuitive way the PT law is encoded in the relation \cite{b2} 
${\bf{\dot{P}}} = -(\bf P.	{\bf{\dot{r}}} ) \bf r$
where the vector $\bf P$ is being transported along a loop $\bf r(T)=\bf r(0)$ with $\bf r(t)$ being the unit radius vector. The PT law requires that $\bf P. \bf r(t)=0$ that is $\bf P$ is not allowed to twist around $\bf r(t)$. In the classic example of the loop on $S_2$, the rotation in {\bf{P}} is converted to a complex phase, the measure of anholonomy, and the closed line integral of the phase is converted to a surface integral via Stokes theorem, leading to the well known result that the anholonomy is given by the solid angle subtended at the centre of the sphere by the area enclosed by the loop. In this work we will study the anholonomy when $\bf P$ is parallel transported along  noncontractible cycles on a torus and come up with some surprising new results. Our results show that there is non-zero  anholonomy for toroidal loops but no anholonomy for poloidal loops. But most interestingly for the more complicated cases of unknots and knots (that contain both toroidal and poloidal loops  \cite{knot,ober}) the anholonomy identically vanishes the reason being that  the torus has both positive and negative curvatures.

 Solution of the  PT equation \cite{b2} will provide locally the explicit structure  of the vector $\bf P$ that is being transported  along each point on the path.   Taking $\bf{u}\mid_\phi$,  the unit tangent vector of the path  at $\phi$,  we will compare the projection of $\bf P$ on $\bf u\mid_{\phi =0}$ before  the cycle, $\bf {P}\mid_{\phi =0}. \bf{u}\mid_{\phi =0}$,  with  the projection of $\bf P$ on $\bf u\mid_{\phi =0}$ after  the cycle, $\bf {P}\mid_{\phi =2\pi p}. \bf{u}\mid_{\phi =0}$. Note that this method may not be totally foolproof since there is a chance of loosing anholonomy information if  $\bf P$  rotates by integer multiples of $2\pi $. We consider a regular torus to be embedded in  three dimensional space.

Instead of using PT law as given in \cite{b2} we use an equivalent form  (see for example \cite{usu}),
\begin{equation}
	\label{5}
	u^iD_iP^j=0,
\end{equation}	
	where $\bf D$ denotes the covariant derivative.  PT does not change length of the vector that is being transported since $u^iD_i\mid {\bf{P}}\mid ^2 =0$.

	The regular torus is parameterized by $\phi, \theta $ (see figure 1),	
\begin{equation}
\label{1}
x_1=(c+acos\theta)cos \phi; ~~x_2=(c+acos\theta)sin \phi;~~x_3=asin\theta 
\end{equation}	
with the metric and non-zero Christoffel connection components,
\begin{equation}
\label{2}
ds^2=(c+acos \theta )^2(d\phi)^2+a^2(d\theta)^2=g_{\phi\phi}(d\phi)^2+g_{\theta\theta}(d\theta)^2,
\end{equation}	
\begin{figure}[htb!]
	{\centerline{\includegraphics[width=6cm, height=6cm] {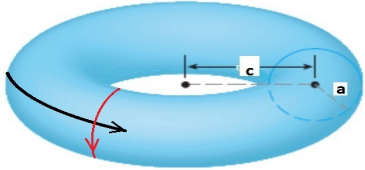}}}
	\caption{Red and black directions refer to poloidal ($\theta $) and toroidal ($\phi $) directions. Parameters $c$ and $a$ for a toroidal surface are shown.} \label{fig1}
\end{figure}
\begin{equation}
\label{3}
\Gamma ^\phi_{\phi\theta}=\Gamma ^\phi_{\theta\phi}=-\frac{asin\theta}{c+acos\theta}~,~~\Gamma ^\theta_{\phi\phi}=-\frac{sin\theta(c+acos\theta)}{a} .
\end{equation}	
\vspace{.1cm}
{\bf{Poloidal cycle:}} 
  $\phi$ remains fixed and only $\theta$ varies ($x^i=\phi_0,\theta$). The unit tangent vector, 
$
{\bf{u}}=\{0,1\}/a\equiv \{0, u^\theta\}.
$ along with 	
$ u^\theta D_\theta P^j=0$ from (\ref{5}) lead to,
\begin{equation}
\label{6i}
\partial_\theta P^\phi +P^\phi \Gamma ^\phi_{\phi\theta}=\partial_\theta P^\phi -\frac{asin\theta}{c+acos\theta}P^\phi =0,~~
\partial_\theta P^\theta  =0.
\end{equation}	
Exploiting the boundary conditions $\theta =0,~P^\theta=\bar P_0^\theta,~ P^\phi=\bar P_0^\phi$ (for a fixed $\phi = \phi_0$) the solutions are,
\begin{equation}
\label{5a}
P^\theta =\bar P_0^\theta,~~P^\phi =\frac{(c+a)\bar P_0^\phi}{c+a cos\theta }.
\end{equation}
Note  that  $\mid {\bf{P}}\mid ^2 $ remains constant:  
$
{\bf{P}}^2\mid_{(\phi_0,\theta=0)}={\bf{P}}^2\mid_{(\phi_0,\theta)}=(c+a)^2(\bar P^\phi_0)^2+a^2(P^\theta_0)^2.
$
The projection of $\bf{P}$ (at beginning and  ending of loop) on the tangent vector (at the starting of the loop) do not change, 
$
{\bf{u.P}}\mid_{(\phi_0,0)}={\bf{u.P}}\mid_{(\phi_0,\theta)} =a\bar P_0^\theta 
$, indicating that  for all poloidal loops there is no anholonomy, the reason being  that the poloidal loops are geodesics of the torus and hence PT along them can not generate anholonomy. But interestingly, later on we will show that this is true even for slanted non-geodesic loops that is when $\phi $ also changes (but not enough to form a toroidal loop as well).

To define enclosed area by  noncontractable loops (in order to apply Stokes theorem) a generalization has been suggested by Hannay \cite{h2} where one has to consider a reference loop in conjunction with the loop in question and join them by a thin neck. A comment about the reference loop is in order. We point out that  on a compact two-dimensional non-simply-connected Riemannian
manifold, each homotopy class of closed loops contains at least one
geodesic, for which the holonomy necessarily vanishes. Hence it is natural to choose the geodesic as the reference loop. In the line integral the contribution of the neck cancels out and one is left with only the contribution of the loop in study. The advantage is that this extended loop structure encloses an area in the conventional way so that Stokes theorem can, in principle, be applied for closed paths with knots (and unknots). We have not pursued this idea  here.

\vspace{.1cm}
{\bf{Toroidal cycle:}} 
$\theta$ is fixed and  only $\phi$ changes and  ($x^i=\phi,\theta_0$) leading to the unit tangent vector 
$
{\bf{u}}=\{1,0\}/(c+acos\theta_0)\equiv \{u^\phi,0\}.
$	
Hence (\ref{5}) simplifies to $ u^\phi D_\phi P^j=0$ leading to the equations,
\begin{equation}
\label{9}
\partial^2_\phi P^\phi +sin^2\theta_0~ P^\phi =0,~~\partial^2_\phi P^\theta +sin^2\theta_0~ P^\theta =0,
\end{equation}	
with solutions and  $\alpha =sin\theta_0 $
\begin{equation}
\label{10}
P^\phi =Acos(\alpha\phi )+Bsin(\alpha\phi );~~P^\theta =Ccos(\alpha\phi )+Dsin(\alpha\phi ).
\end{equation}	
 Incidentally this is similar to the  result that one gets for loops $S_2$.  Introducing the boundary conditions, $\phi =0$, $P^i=(P^\phi_0,P^\theta_0 )$ the complete solutions are
\begin{equation}
\label{11}
P^\phi =P^\phi_0cos(\alpha\phi )+\frac{aP^\theta_0}{c+acos\theta_0}sin(\alpha\phi ),~
P^\theta =P^\theta _0cos(\alpha\phi ) -\frac{(c+acos\theta_0)P^\phi_0}{a}sin(\alpha\phi ).
\end{equation}
 $
{\bf{P}}^2\mid_{(0,\theta_0)}={\bf{P}}^2\mid_{(\phi,\theta_0)}=(c+acos\theta_0)^2(P^\phi_0)^2+a^2(P^\theta_0)^2$ remains unchanged.

Let us now consider the behavior of the projection,
\begin{equation}
\label{25}
{\bf{u.P}}=g_{\phi\phi}u^\phi  P^\phi +g_{\theta\theta}u^\theta P^\theta =(c+acos\theta_0)P^\phi_0 cos(\alpha\phi ) +aP^\theta_0 sin(\alpha\phi ),
\end{equation}
at different stages of rotation with its reference value at the beginning $(\phi =0)$,
\begin{equation}
\label{26bb}
(u.P)\mid_{(0, \theta_0)}=(c+acos\theta_0)P^\phi_0 .
\end{equation}
After a full rotation the projection is
\begin{equation}
\label{27}
(u.P)\mid_{(2\pi, \theta_0)}=(c+acos\theta_0)P^\phi_0 cos(2\pi\alpha)+aP^\theta_0 sin(2\pi\alpha);~~\alpha =sin\theta_0,
\end{equation}
which is clearly different from (\ref{26bb}). To indicate anholonomy let us define a parameter $\Sigma (\theta_0)$ that is the ratio of the projections at the beginning $(\phi=0)$ and ending $(\phi=2\pi)$ of a complete cycle:
\begin{equation}
\label{25a  }
\Sigma (\theta_0)=\frac{({\bf{u.P}})\mid_{(2\pi,\theta_0)}}{({\bf{u.P}})\mid_{(0, \theta_0)}}=cos (2\pi \alpha ) +\frac{aP_0^\theta sin(2\pi\alpha)}{(c+acos\theta_0)P_0^\phi}.
\end{equation}
Figure 3 shows the dependence of $\Sigma (n)$ on $\theta_0$,

\begin{equation}
\label{anhol}
\Sigma (n)=cos (2\pi \alpha ) +\frac{(n+1)sin(2\pi\alpha)}{n+cos\theta_0}
\end{equation}
where we have scaled $c=na$ and chosen $a=1,~P_0^\phi =1/((n+1)\sqrt 2),~P_0^\theta =1/\sqrt 2)$ so that for any $n$ at $\theta_0=0$, ${\bf{P}}^2=1$. {\it{Deviation of $\Sigma (n)$ from unity indicates anholonomy}}. We can see that the position of the toroidal loop, $\theta_0$, as well as dimensions of the torus $n$, drastically change the anholonomy. $n\ge 1$ corresponds to the standard "ring" torus (blue line in Figure 2) whereas surfaces with $n\le 1$ leads to degenerate configurations. In particular, for $n=1$ (green line in Figure 2) the inner radius of the torus goes to zero, and origin has a limiting behavior without a tangent plane. For  $n= 1$ $n\le 1$ the torus self-intersects, (here chosen $n=0.5$) and we have a "spindle" tori (red line in Figure 2). (See \cite{jan} for some details.) The main point we wish to emphasize  is that whenever $n\le 1$ the anholonomy  $\Sigma (n)$ will diverge at some points because of the denominator $n+cos\theta_0$ in (\ref{anhol}). This is seen in the red and green lines  in Figure 2. The blue line in Figure 2 with $n=3$ gives the proper representation of anholonomy in ring torus.
\begin{figure}[htb!]
	{\centerline{\includegraphics[width=6cm, height=4cm] {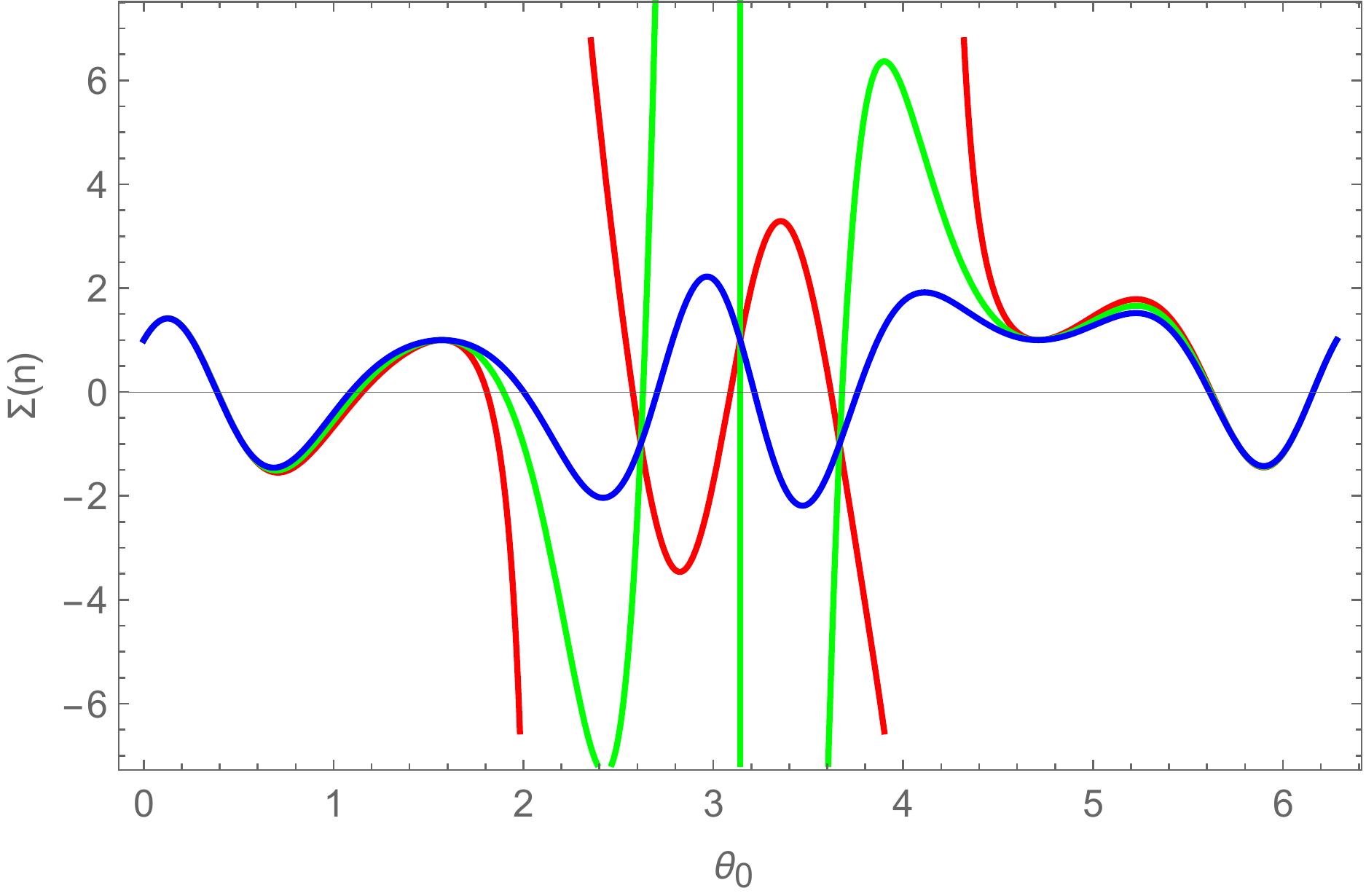}}}
	\caption{$\Sigma (n) $ vs. $\theta_0$ graph for $a=1$ and $n=0.5,1,3$: red, green and blue lines respectively} \label{fig2}
\end{figure}

 Some special values of $\theta_0$ will be of interest. $\theta_0=0, \alpha =0$ and  $\theta_0=\pi, \alpha =0$ correspond to the largest (or outermost) and smallest (or innermost) toroidal cycles, which incidentally are geodesics of the torus \cite{jan}. That these two toroidal loops as well as all poloidal loops (or meridians) are geodesics is understandable from noting that the torus has a reflection symmetry along these lines.  We find,
 absence of anholonomy along these paths.

 Again, $\theta_0=\pi/2, \alpha =1$ and $\theta_0=3\pi/2, \alpha =-1$ correspond to the uppermost and lowermost circles having radius $c$, for which, once again, the projections do not change indicating no anholonomy,

Moving slightly away from these special $\theta_0$ values, for small  $\pm\theta_0$, we find $ \alpha=sin(\pm\theta_0 )\approx \pm\theta_0,~cos  (\pm\theta_0)\approx 1$
\begin{equation}
\label{27a}
(u.P)\mid_{(0, \pm\theta_0)}=(c+a)P^\phi_0,~~(u.P)\mid_{(2\pi,\pm  \theta_0)}=(c+a)P^\phi_0 cos(2\pi\theta_0) \pm aP_0^\theta sin(2\pi\theta_0),
\end{equation}
so that 
\begin{equation}
\label{25ab}
\Sigma (\pm\theta_0)=\frac{(c+a)P^\phi_0 cos(2\pi\theta_0) \pm aP_0^\theta sin(2\pi\theta_0)}{(c+a)P^\phi_0}.
\end{equation}
In regions close to $\theta_0 =\pi$, $\alpha =sin(\pi \pm\theta_0 )\approx \mp\theta_0, ~cos(\pi +\theta_0 )\approx -1$:
\begin{equation}
\label{25abc}
\Sigma (\pi\pm\theta_0)=\frac{(c-a)P^\phi_0 cos(2\pi\theta_0) \mp aP_0^\theta sin(2\pi\theta_0)}{(c-a)P^\phi_0}.
\end{equation}

Now for regions close to $\theta_0 =\pi/2$ and $\theta_0 =3\pi/2$ we note that to first order the anholonomy vanishes. 
Later on we will comment on  slanted loops that is when $\theta $ also changes but not enough to form a poloidal loop as well.

\vspace{.2cm}
{\bf{Knots and Unknots - combination of Poloidal and Toroidal loops:}} \\
Knot and unknots are closed curves that wrap around the surface of a torus (see \cite{ober} for a detailed discussion), typically symbolized  by $T_{p,q}$ with $p,q$ integers, meaning that the curve wraps $p$ times and $q$ times respectively along the toroidal and poloidal directions respectively. For instance $T_{2,3}$ represents the well known trefoil knot.  If $p,q$ are co-prime integers we have a knot and if either $p=1$ or $q=1$ we have a multiply coiled curve that is an unknot. $T_{p,q}$ and $T_{q,p}$ are topologically equivalent which means that one can be smoothly deformed in to the other but their local geometric properties are different. The winding number  $\omega =q/p$ is a measure of the complexity of the closed curve. 

We parameterize the angle variables by $\phi \rightarrow \phi,~\theta \rightarrow \omega\phi $ with $2\pi p \geq\phi \geq 0$, so that with $x^i=\{\phi,\theta\}=\{\phi,\omega\phi\}$ we get   the unit tangent vector,
$
{\bf{u}}=\{1,\omega \}((c+acos\omega\phi)^2+a^2\omega^2)^{-1/2}\equiv \{u^\phi,u^\theta\}.
$	
 Defining
 $c+acos\omega\phi =z,~P'=dP/dz$ the decoupled equations,
\begin{equation}
(P^\theta )''+\frac{1}{a^2\omega^2}P^\theta =0,~~(P^\theta )'-\frac{z}{\omega a^2}P^\phi =0.
\label{a3}
\end{equation}
have solutions:
\begin{equation}
\label{a4}
P^\theta =Acos(\frac{z}{a\omega}) +Bsin(\frac{z}{a\omega}),~~P^\phi =\frac{a}{z} (-Asin(\frac{z}{a\omega})+ Bcos(\frac{z}{a\omega})).
\end{equation}
From the boundary conditions $P^\theta =P^\theta _0,~P^\phi =P^\phi _0 $ at $\phi =0,~z=c+a=a(n+1)$ and with a further scaling $c=na$, the constants $A,B$, are
\begin{equation}
B=(n+1)P^\phi _0 cos \frac{n+1}{\omega}+P^\theta _0 sin \frac{n+1}{\omega},~~A=\frac{P^\theta _0-Bsin \frac{n+1}{\omega}}{cos \frac{n+1}{\omega}}.
\label{a5}
\end{equation}
It is simple to check that $
P^2=a^2((n+1)^2 (P^\phi _0)^2+(P^\theta _0) ^2)
$ 
remains invariant under PT.  As before we compute the projection
\begin{equation}
\label{a7}
({\bf{u.P}})\mid _{(\phi =0)}=\frac{a}{{\sqrt{(n+1)^2+\omega ^2}}}(P^\phi _0+\omega P^\theta _0)
\end{equation}
Now notice that, after a complete cycle, $\phi =2\pi p$, the variable $z=a(n+cos (\frac{q}{p}\phi ))$ returns  to its value $z=a(n+1)$ at $\phi =0$ and so there will not be any change in the projection ${\bf{u.P}}$ after a complete cycle. This means that for any kind of complex loop, be it knot or unknot, there will not be any anholonomy. An intuitive explanation of this surprising fact is that the torus has positive curvature at the outside and negative curvature at the inside. Since the rotation of a vector after parallel transport  has opposite signatures for positive and negative curvatures the net angle variation cancels out if the path completes one or more poloidal loops. We have already seen this principle at work in case of simple poloidal loops.

Now that we have a general result of PT for an arbitrary complex loop, we can also generalize our previous examples of  poloidal loops and toroidal loops by making them slanted that is where both $\theta, \phi $ angles vary but the complete loop is performed in either poloidal or toroidal directions only. This means that for a slanted poloidal loop, $q$ is an integer but $p\le 1$. Then again $z=a(n+cos \omega 2\pi p)=a(n+cos \omega 2\pi q)=  a(n+1)$ so that there will be no anholonomy. On the other hand if the toroidal loop is slanted then $p$ will be an integer but $q\le 1$ in which case anholonomy will be non-vanishing. 

We stress that the vanishing anholonomy result (in a curved surface) for the poloidal loop is novel and unexpected whereas toroidal loop results are of expected nature.
\vskip .3cm
{\bf{Hannay angle for noncontractible loops}}

One of the problems considered by Hannay \cite{h1} is a particle sliding frictionlessly on a planar hoop of arbitrary shape where the hoop itself is also rotating about an axis perpendicular to the plane of the hoop. The positions of the particle on the hoop, with and without rotation of the hoop itself, differ by an angle, Hannay angle, that is a geometrical object depending on the area enclosed by the hoop. Stokes theorem plays an important (although not essential) role in the demonstration for an arbitrarily shaped hoop. We will generalize the system to  a particle  moving on a regular torus with the torus itself revolving in space. The extension from the hoop, (with first homotopy group $\pi_1 (hoop)\approx Z^1$), to torus, ( $\pi_1 (torus)\approx Z^1\times Z^1$),  is non-trivial due to the   nature of noncontractable loops. We verify specific cases of a very general statement by Hannay \cite{h2}, "(Hannay) angle, or rather the associated shift, for a general rigid rotation of a general loop (planar or not) in space about a fixed axis is ($4 \pi$ times) the projected area of the loop onto a plane perpendicular to the rotation axis divided by the length of the loop".  

Hannay's formulation \cite{h1} starts from the generic relation $d/dt \mid_{space} =d/dt \mid_{body}+{\bf{\Omega}}\times $ that connects time derivatives of a vector between  static frame  and a frame rotating with ${\bf{\Omega }}$. When applied to the position ${\bf{r}}$ of a particle this yields
$
\bf p\mid_{space} =\bf p \mid_{body}+\bf\Omega \times \bf r$, connecting the momenta for a unit mass particle. The adiabetic invariant quantity is the action (line integral around the loop)
\begin{equation}
\label{h2}
I=\frac{1}{2\pi}\oint \bf p.d\bf r =\frac{1}{2\pi}\oint (\dot{\bf r}+\bf\Omega \times \bf r).d\bf r. 
\end{equation}
Exploiting Stokes  theorem (when applicable) the second term can be written as $2{\bf{\Omega }}.{\bf{A}}/L $ where $A$ and $L$ are the area enclosed by  and length of the loop respectively. This term leads to the shift in angle for the particle motion  $\Delta \theta =-8\pi^2A/L^2 $ that is the Hannay angle \cite{h1}. However in the present work we will stick to the  line integral prescription since we focus on noncontractible loops.

Berry \cite{b3} has shown that there are alternative ways of recovering the Hannay angle and one such way is considering Newtonian dynamics with Euler pseudo forces that operate in the non-inertial rotating (torus) frame. It is strightforward to establish,
\begin{equation}
\label{h3}
\ddot{s}(t)=\bf t. (-\bf\Omega \times (\bf\Omega\times \bf r)-\dot{\bf\Omega }\times r )=\frac{d{\bf r} (s(t))}{ds}.(\Omega^2{\bf r}-({\bf \Omega}.{\bf r}){\bf \Omega}-\dot{\bf\Omega }\times r )
\end{equation}
where $s(t)$ is an arc length measured from a specific point on the loop and $\bf{t}$ is the unit tangent vector. Note that the Coreolis force $\sim~\bf\Omega \times \dot{\bf r}$ is absent since the particle is forced to move in a fixed path and this force is normal to the path. After  two integrations the implicit solution is obtained as
\begin{equation}
\label{bb1}
s(t)=s_0+p_0t +\int_0^t dt' (t-t')F(t',s)
\end{equation}
where we have clubbed the RHS of (\ref{h3}) as $F(t)$. Finally assuming that $\Omega $ and $\dot{\Omega} $ are small that is, the particle completes the circuit many times in time $T$ during which the loop revolves only once, the $s$-dependence can be averaged out to yield  
\begin{equation}
\label{bb20}
s(T)=s_0+p_0T +\int_0^Tdt'~[\frac{1}{L}\int_0^L ds  (t-t')F(t',s)].
\end{equation}
The $s$ or equivalently angle averaging corresponds to the adiabetic principle \cite{arnold}. Contribution from the last term in RHS of (\ref{bb20}) with proper scaling reproduces
 the anholonomy or Hannay angle. Below we will use both (\ref{h2}) and (\ref{bb20}) to compute  Hannay angles for different noncontractible loops on torus.

{\bf{Toroidal loops:}} For toroidal loops with  fixed $\theta =\theta_0 $ and
\begin{equation}
\label{a1}
{\bf{r}}\equiv \{(c+a~cos\theta_0)cos \phi, ~(c+a~cos\theta_0)sin \phi,~a~sin\theta_0\};~~ {\bf{\Omega}}\equiv \{\Omega_1, ~\Omega_2,~\Omega_3\},
\end{equation}	
we find
\begin{equation}
\label{h5}
\oint ({\bf \Omega}\times {\bf r}).d{\bf r}=\int _0^{2\pi }d\phi \chi [-\Omega_2a~sin\theta_0 sin\phi -\Omega_1 a~ sin\theta_0 cos\phi +\Omega_3 (c+a~cos\theta_0)]$$$$=2\pi (c+a~cos\theta_0) ^2\Omega_3 \equiv 2A\Omega_3
\end{equation} 
where $A$ denotes area of the loop - a circle with radius $c+a~cos\theta_0$.
The last form agrees with the structure obtained by Hannay \cite{h1} for a circular hoop since qualitatively toroidal loops with fixed $\theta_0$ is same as a circular hoop. Due to anholonomy, the (spatial) average speed is greater by $2A\Omega_3/L$ and for small $\Omega$ one equates spatial average to time average. Thus over one revolution of the loop in $\Omega_3$, $\int_0^T \Omega_3dt=2\pi$ and the extra distance traversed is $-4\pi A/L$ or the extra angle (Hannay angle)  is obtained by multiplying by $2\pi/L$ to get 
\begin{equation}
\label{hh12}
\Delta \Theta =-\frac{8\pi^2 A}{L^2}.
\end{equation}
For circular loop this is simply $-2\pi$. The fact the Hannay angle depends only on $\Omega_3$ agrees with the general statement of Hannay \cite{h2} since the toroidal loop  in horizontal plane has zero projection on planes perpendicular to $\Omega_1$ and $\Omega_2$.

Now from Berry's method, with for (\ref{a1}) with 
${\bf t}=\{-sin\phi,~cos\phi ,~0\}$
we get
\begin{equation}
\label{b1}
\ddot {s}=\frac{1}{2}(c+acos\theta_0) (\Omega_2^2-\Omega_1^2)sin(2\phi) +\Omega_1\Omega_2 (c+acos\theta_0) cos(2\theta )+a\Omega_3sin\theta_0(\Omega_2cos\phi -\Omega_1 sin\phi )$$$$+asin\theta_0(\dot {\Omega_2}sin\phi +\dot {\Omega_1}cos\phi)-(c+acos\theta_0) \dot{\Omega_3}
\end{equation}
Only the last term with $\dot{\Omega_3}$ will survive the loop averaging to yield 
$$\Delta s= -\int_0^T dt' (T-t')(c+acos\theta_0) \dot{\Omega_3}. $$   
Using $\int_0^T dt' (T-t')\dot{\Omega_3}=2\pi$ we recover
\begin{equation}
\label{bb2}
\Delta s  =-2\pi (c+a~cos\theta_0).
\end{equation}
Now $\Delta\Theta=(2\pi/L)\Delta s=-2\pi $.
Note that this result is a straightforward extension of the hoop result since the closed path considered here for fixed $\theta=\theta_0$ is similar to the motion along a hoop.

{\bf{Poloidal loops:}} Coming to poloidal loops with fixed $\phi=\phi_0$  we get
\begin{equation}
\label{h6}
\oint ({\bf \Omega}\times {\bf r}).d{\bf r}=\int_0^{2\pi } d\theta [-\Omega _2(a~cos\phi_0 -c~cos\theta cos\phi_0)+\Omega_1(a~ sin\phi_0 +c~ cos\theta sin\phi_0)]$$$$=2\pi a^2(\Omega_1sin\phi_0-\Omega_2cos\phi_0) \equiv 2\bar A(\Omega_1sin\phi_0-\Omega_2cos\phi_0).
\end{equation}
 Once again the general statement of Hannay \cite{h2} is satisfied since  the poloidal loop in vertical  plane has zero projection on plane perpendicular to $\Omega_3$ and Hannay angle depends only on $\Omega_1$ and $\Omega_2$.  The axial symmetry is broken making the result $\phi_0$-dependent. Consider only $\Omega_1$ to be non-zero in which case anholonomy will vanish for $\phi_0=0,\pi $ because these two loops are on the axis of rotation of the loop and so do not experience the rotation effect. If either $\Omega_1$ or $\Omega_2$ is non-zero the Hannay angle follows similar arguments but if both $\Omega_1$ and $\Omega_2$ are non-vanishing the above argument needs to be generalized to $\int_0^{\tilde T} \Omega_1dt=2\pi n_1,~\int_0^{\tilde T} \Omega_2dt=2\pi n_2$  with $n_1,n_2$ being smallest integers satisfying the condition that the torus returns to itself after revolution. Hence the Hannay angle is
\begin{equation}
\label{hh1}
\Delta \Theta =-\frac{8\pi^2\bar A}{L^2} (n_1 sin\phi_0-n_2 cos\phi_0).
\end{equation}
Notice that Hannay angle disappears for $n_1 sin\phi_0=n_2 cos\phi_0$.
Hannay angle for poloidal cycle, an example of noncontractible loop of a different type, is a new  and one of our major results.

Let us recover the same result following Berry's analysis where ${\bf t}\equiv \{-cos\phi_0 sin\theta ,~sin\phi_0 sin\theta, ~cos\theta \}$
 leads to
\begin{equation}
\label{b2}
\ddot{s}=-(\Omega_1^2cos^2\phi_0 +\Omega_2^2sin^2\phi_0+\Omega_1\Omega_2sin(2\phi_0)) (c+a~cos\theta )sin\theta $$$$+\frac{1}{2}a\Omega_3^2 sin(2\theta)+(\Omega_1 cos \phi_0+\Omega_2 sin \phi_0) \Omega_3(c~cos\theta+a~cos(2\theta ))-c~{\bf{\Omega}}^2sin\theta$$$$ -(\dot{\Omega_1} sin \phi_0-\dot{\Omega_2} cos \phi_0)(a+c~cos\theta ),
\end{equation}
and only the $a$-term in the last expression survives the $\theta$-integration and same procedure as discussed in the toroidal case will yield (\ref{hh1}).\\
{\bf{$p,q$ Torus knot:}}
We parameterize the angle variables  by $\phi \rightarrow p\phi,~\theta \rightarrow q\phi $ with $2\pi  \geq\phi \geq 0$, that is same as before (see above (\ref{a3})) with $\omega = q/p$. From the position  vector,	
\begin{equation}
\label{k1}
{\bf{r}}\equiv \{(c+acos q\phi)cos p\phi; ~~(c+acos q\phi)sin p\phi;~~x_3=asin q\phi \}
\end{equation}	
we compute
\begin{equation}
\label{k2}
\oint{\bf{\Omega}}\times {\bf{r}}.d{\bf{r}} =\int_0^{2\pi}d\phi~ p\Omega_3(c+acosq\phi)^2=p(2c^2+a^2)\pi \Omega_3.
\end{equation}
The total length of the $p,q$-knot is given by
\begin{equation}
\label{kk2}
L=\int _ 0^ {2\pi} d\phi =[(\frac{dx_1}{d\phi})^2+(\frac{dx_2}{d\phi})^2+(\frac{dx_3}{d\phi})^2]^{1/2}=a\int_0^{2\pi}d\phi [q^2+p^2(n+cosq\phi)^2]^{1/2}
\end{equation}
and an approximate form can be obtained from the average of the maximum and minimum value of the expression:
\begin{equation}
L\approx\pi a ([q^2+p^2(n+1)^2]^{1/2}+[q^2+p^2(n-1)^2]^{1/2})
\label{k22}
\end{equation}
Hence a form of the Hannay angle is,
\begin{equation}
\label{hhh1}
\Delta \Theta =-\frac{2\pi^2}{L^2}p(2c^2+a^2)\pi .
\end{equation}
The computation of the same from Berry's framework is computationally more involved due to the complicated nature of the unit tangent vector,
\begin{equation}
\label{k4}
{\bf{t}}\equiv \frac{1}{[q^2a^2+p^2(c+acos q\phi )^2]^{1/2}]}$$$$\{(-aq~sinq\phi~cosp\phi -p((c+acos q\phi )sinp\phi),~(-q~sinq\phi~sin p\phi +p((c+acos q\phi )cosp\phi),~q~acos q\phi\}
\end{equation}
For the special case  ${\bf{\Omega}}=\Omega_3\hat k$, (since we have seen from (\ref{k2}) that only $\Omega_3$ contributes),
\begin{equation}
\label{k5}
\ddot s=-\frac{p(c+acos q\phi )^2}{[q^2a^2+p^2(c+acos q\phi )^2]^{1/2}}\dot{\Omega_3}-\frac{(c+acosq\phi )aqsinq\phi }{{[q^2a^2+p^2(c+acos q\phi )^2]^{1/2}}}\Omega_3^2
\end{equation}
and it is clear that only the first term ($\sim \dot{\Omega_3}$) will survive the loop integration. Comparing with (\ref{k2}) the structures are obviously similar but we do not pursue with this computation any further and remain satisfied with (\ref{hh1}) as the Hanny angle for the $p,q$-knot.
\vskip .3cm

\vskip .3cm
{\bf{Conclusion and future directions}}\\
To summarize we have computed anholonomy and Hannay angle  for noncontractible  loops on a torus, in the form of toroidal and poloidal cycles and knots (containing both these forms of loops). These type of loops have not been studied before in the present context. An interesting form of  anholonomy cancellation phenomenon for closed paths that include poloidal loops is revealed. Although the attached vector rotates locally under parallel transport, the final anholonomy cancels out since the particle visits zones having positive and negative curvatures. 

We have explicitly verified a general observation by Hannay concerning the relation between Hannay angle, the loop itself and the rotation axis of the loop. The Hanny angles for particle moving on  noncontractible cycles when the torus itself is revolving arev computed. We have used two distinct frameworks suggested by Hannay \cite{h1} and Berry \cite{b2} and the results agree (albeit for knots where we have indicated the agreement in leading order). A new and interesting result is that there is more structure in the Hannay angle expression for poloidal loop, not anticipated earlier, although the toroidal loop result is  essentially similar to Hannay's result for a planar hoop \cite{h1}.

Let us list some of the interesting  future directions of work.\\
(i) A problem of immediate interest is to compute the Berry phase associated with the corresponding quantum problem that we plan to include in a publication in near future.\\
(ii) It will be very interesting to introduce a magnetic field and study the revolving torus along with the magnetic field's effect on Berry phase and Hannay angle. Quantum mechanics of particle on a static torus has appeared in \cite{mag} and also in \cite{sree,g1} in a semi-classical framework.\\
(iii)  A natural and non-trivial generalization would be to consider a generic form of torus that is, a deformed torus, compared to the canonical (regular) form of torus studied here. This will indicate how general (or topological) our results are. A possible way to do it is to work with a toroidal coordinate system and then try to see effects of small variations about fixed $c$ (azimuthal radius) and $a$ (ring radius). The same goes for generic nature of loops as well. Another interesting aspect is to generalize the present analysis to higher genus surfaces.\\
(iv) As a practical realization in the classical context one can think about a particle possessing a constant vector fixed on it such as a spinning top or a gyroscope, (though in case of top the spin vector is radial whereas we  have considered the vector to have vanishing radial component), with the particle sliding over a torus. In an optical setup one can think of light moving in a  torus knot loop of  optical fibre \cite{h2}. It is an open problem for experimentalists to construct time dependent magnetic fields that can simulate noncontractible cycles in parameter space.

\vspace{.1cm}
{\bf{Acknowledgments:}} It is a pleasure to thank Professor John Hannay for many helpful correspondences and concrete suggestions. I am grateful to Professor Michael Berry for helpful comments. Also I thank  the referee for constructive comments.

\end{document}